\documentclass[letterpaper,12pt]{amsart}
\usepackage[english]{babel}
\usepackage{graphicx}
\usepackage{amsmath}
\usepackage{listings}
\usepackage{parskip}
\usepackage{gensymb}
\usepackage[colorlinks]{hyperref}
\usepackage{mathrsfs}
\usepackage{csquotes}
\usepackage{float}
\usepackage[labelfont=bf]{caption}
\usepackage{rotating}
\usepackage{float}
\restylefloat{table}
\usepackage{subfigure}
\usepackage{color, colortbl}
\definecolor{LRed}{rgb}{1,.8,.8}
\definecolor{MRed}{rgb}{1,.6,.6}
\definecolor{HRed}{rgb}{1,.2,.2}
\usepackage{color}

\usepackage{pdfpages}

\title[Approximate 1-D Models for Transport in Ducts]{Approximate One-Dimensional Models for Monoenergetic Neutral Particle Transport in Ducts with Wall Migration}
\author{Arnulfo Gonzalez}  \author{Ryan G.\ McClarren}\email{rgm@tamu.edu}\address{Dept.\ of Nuclear Engineering\\ Texas A\&M University\\ College Station, TX, 77843-3133\\ USA}
\begin{document}
\maketitle

\begin{abstract}
The problem of monoenergetic neutral particle transport in a duct, where  particles travel inside the duct walls, is treated using an approximate one-dimensional model. The one-dimensional model uses three-basis functions, as part of a previously derived weighted-residual procedure, to account for the geometry of particle transport in a duct system (where particle migration into the walls is not considered).  Our model introduces two stochastic parameters to account for particle-wall interactions: an albedo approximation yielding the fraction of particles that return to the duct after striking the walls, and a mean-distance travelled in the walls transverse to the duct by particles that re-enter the duct. Our model produces a set of three transport equations with a non-local scattering kernel. We solve these equations using discrete ordinates with source iteration. Numerical results for the reflection and transmission probabilities of neutron transport in ducts of circular cross section are compared to Monte Carlo results computed using the MCNP code. 
\end{abstract}

{\section{ { Introduction}}}
There are several applications that involve the transport of particles through channels where there is a weakly interacting medium of constant cross-section surrounded by a strongly interacting medium. These channels are often called ducts or pipes because they often appear in situations where there is a low-density medium (even vacuum) surrounded by a dense material. Such situations arise in charged-particle transport \cite{Prinja:1984tr}, radiation shielding \cite{schaeffer1973reactor}, and acoustics \cite{Jing:2010cl,Jing:2010hq,Visentin:2012cm}.

Prinja and Pomraning\cite{Prinja:1984tr} first developed a one-dimensional model for neutral particle transport in ducts using geometric arguments. In that work, the distance traveled between wall collisions was interpreted as an angle-dependent cross-section. Using the method of weighted residuals, Larsen and colleagues in two separate papers \cite{Larsen:1984ue,Larsen:1983uh} developed a rigorous mathematical formulation of the same one-dimension model, and demonstrated its equivalence to the lowest order approximation of a chain of approximations. The weighted residual method is based on using $N$ basis functions to approximate the azimuthal and transverse components of the particle distribution function in the duct, and $N$ weight functions to transform the original domain to have a single spatial and single angular variable. 

The weighted-residual approach with a single basis function ($N=1$) gives the same result as the geometric approach originally promulgated by Prinja and Pomraning. Adding a second basis function was shown to improve the accuracy of the 1-D models significantly \cite{Larsen:1983uh}. Later, Garcia, Ono, and Veira \cite{Garcia:2000bw} developed a third basis function. Each additional basis function included in the 1-D model leads to improved accuracy compared with high-fidelity, 3-D calculations . For the $N=1$ model, applied to circular ducts with length to radius ratios between 0.1 and 10,  Ref.\ \cite{Garcia:2000bw} reports a maximum percent deviation of 32.8 \% for reflection probability, and over 300\% in the transmission probability. For the $N=2$ and $N=3$ cases, the maximum percent deviation reported for  reflection probabilities are 9\% and 5\%, and  1.3\% and 0.27 \% for the transmission probabilities. 
Besides these works developing more accurate models, there has also been research into efficient solution techniques for these one-dimensional models \cite{Garcia:2015hw,Ganapol:2015fk,Garcia:2015cs,Garcia:2014iv,Garcia:2013vj,Barichello:2011hb,Garcia:1999jh}.

Each of the approximate one-dimensional models was based on the assumption that particles do not penetrate the walls---that is, they are reflected or absorbed; as such, the material of the walls was not taken into account. Garcia, Ono, and Veira \cite{Garcia:2003fl} expanded on the one-dimensional models by using multigroup albedo approximations to describe particle reflection corresponding to specific wall material (i.e. iron and concrete). The multigroup albedo model was compared to MCNP simulations of neutrons streaming in an evacuated circular duct of fixed length and wall thickness of 100 cm and 20 cm, with and without wall migration. A comparison of the results indicate fairly uniform agreement between the $N=3$ model and the modified MCNP simulations without wall migration; however there is very poor agreement once the MCNP simulations account for wall migration. For circular concrete ducts, ranging in radius from 8 cm to 50 cm, Garcia reports a maximum error of 170\% for reflection probability, while for iron ducts of the same dimensions, he reports a maximum error of approximately 116\%. The transmission probabilities in identical concrete and iron systems give maximum errors of approximately 22\% and 28\%.  

Previous results indicate a need for the development of a one-dimensional model, for two and three basis functions, which can properly account for wall migration. When a particle strikes the wall and is reflected, the albedo model merely supposes that it will reappear at the point of incidence; this is an inadequate approximation given that particles will migrate in the walls. Garcia, Ono, and Veira \cite{Garcia:2003fl} report that in their model up to 50\% of the particles reflected by the wall, just at the duct's edge, will leave the system through the duct entrance without a second interaction. The early exit of particles leads to a vast overestimation of the reflection probability, and underestimation of the transmission probability. 
Prinja \cite{Prinja:1996kl} presented a possible solution to this shortcoming. He derived a 1-D model with a single basis function that accounted for particles moving in the walls via a non-local scattering kernel. He then solved for the reflection probability of a semi-infinite duct using a Weiner-Hopf technique. In this work, we extend Prinja's model by applying it to a three-basis function model, solve problems involving finite ducts, and outline a procedure for calculating the needed model parameters for different wall materials.

\section{ { Model Formulation}}

In this section we introduce the notation necessary to develop our model. Our notation follows that of \cite{Larsen:1983uh}. The duct is assumed to be evacuated (that is there are no particle interactions inside the duct). Monoenergetic particles are introduced into the duct through an open end and stream until striking the inner duct walls. Upon wall collision, the particles are either scattered into the interior of the duct, according to a probability $c$. As such, particles are strictly removed from the system by wall absorption or by streaming out of the duct ends. 

The model assumes a duct parallel to the $z$-axis, with position coordinates ($x,y,z$). The duct has a length $ 0 \leq z \leq Z$, and a cross-section which can be described--- independent of z--- by the function $h(x,y)$,
\begin{subequations}
	\begin{align}
	R: h(x,y) < 0, \\
	\partial R: h(x,y) = 0 ,
	\end{align}
\end{subequations} 
where $R$ defines the duct interior, and $\partial R$ defines the duct's interior wall. For example, a circular duct with radius $\rho$ is described by  
\begin{subequations}
	\begin{align}
	R: \rho^{2} - x^{2} + y^{2} < 0, \\
	\partial R: \rho^{2} - x^{2} + y^{2} = 0.
	\end{align}
\end{subequations}
 The cross-sectional area and duct perimeter are expressed as \[A = \int_{R} \,dx \,dy, \qquad  L = \int_{\partial R} \,ds,\] where $ds$ is the arc length element. Particles in the duct stream with direction $\vec{\Omega}$, which is defined in terms of $\mu \in[-1,1]$ (i.e. the cosine of the polar angle with respect to the z-axis) as well as a corresponding azimuthal angle $\phi \in[0,2\pi]$. The direction vector is defined as
$\vec{\Omega} = ( \sqrt{1 - \mu^{2}}\cos\phi,  \sqrt{1 - \mu^{2}}\sin\phi, \mu)$.

The steady-state transport equation for an evacuated duct 
is expressed as
\begin{equation}
\vec{\Omega} \cdot \nabla \Psi(\vec{r},\vec{\Omega}) = 0.
\label{transport} 
\end{equation}
 Eq.~(\ref{transport}) assumes a steady-state, monoenergetic system without collisions between particles in an evacuated duct, and no external source. The boundary conditions at the ends of a duct of length $Z$ are defined as the prescribed incident fluxes,
\begin{subequations}
	\begin{align}
	\Psi(x,y,0,\vec{\Omega}) = f(x,y,\vec{\Omega}),\: h < 0 ,\, \mu > 0, \\
	\Psi(x,y,Z,\vec{\Omega}) = g(x,y,\vec{\Omega}),\: h < 0 ,\, \mu < 0,
	\end{align}
\end{subequations}
where $f(\mu)$ and $g(\mu)$ are well defined functions. The boundary conditions which describe partial isotropic reflection on the inner wall of the duct are expressed as
\begin{multline}
-\vec{\Omega} \cdot \vec{n} \ \ \Psi(\vec{r},\vec{\Omega}) = \int_{\vec{\Omega}' \cdot \vec{n} \ > \ 0 } p(\vec{r},
\vec{\Omega}' \rightarrow \vec{\Omega}) \Psi(\vec{r},\vec{\Omega})  \partial \vec{\Omega}' \\
h = 0, \ \vec{\Omega} \cdot \vec{n} < 0, 
\label{boundary}
\end{multline}
\begin{equation}
p(x,\vec{\Omega}' \rightarrow \vec{\Omega}) = \frac{-c}{\pi} (\vec{\Omega} \cdot \vec{n}) (\vec{\Omega}' \cdot \vec{n}),
\end{equation} 
where $\vec{n}$ is the unit outward normal at $\vec{r}$, and integration of Eq.~(\ref{boundary}) over $\vec{\Omega} \cdot \vec{n}$  confirms $c$ as the probability of wall reflection \cite{Larsen:1983uh}. Moreover, Eq.~(\ref{boundary}) indicates that for particles incident upon the walls, with incoming direction $\vec{\Omega'}$, the angle of reflection $\vec{\Omega}$ will point towards the interior of the duct. 



The 3-D duct problem is reduced to a 1-D problem using  the method of weighted residuals, where $\Psi$ is approximated as $\psi$,
\vspace{0.5cm}
\begin{equation}
\Psi(x,y,z,\mu,\phi) = \sum_{j=1}^{N} \alpha_{j}(x,y,\phi) \psi_{j}(z,\mu),
\label{weights}
\end{equation}
where the $\alpha_{j}$ are basis functions, and $\psi_{j}$ are expansion functions. Using Eq.~(\ref{weights}) as an approximation for $\Psi$ yields error terms. As such, the method of weighted residuals requires the error terms to be orthogonal to particular weight functions $\beta_{i }(x,y,\phi)$, for $ 1 \leq i \leq N$. 
The basis and weight functions are chosen such that they satisfy 
\begin{equation}\label{eq:normal}
\frac{1}{2\pi A} \int_{R}\int_{0}^{2\pi}  \beta_{i}(x,y,\phi) \  \alpha_{j}(x,y,\phi) \,  d\phi \, d x \, d y = \delta_{ij} , \ \   i,j = 1,2,...,N. 
\end{equation}

As discussed in \cite{Larsen:1983uh}, given prescribed basis and weight functions, the general matrix form of the transport equation with $N$ basis functions, can be expressed as  
\begin{multline}
\mu \frac{\partial \vec{\psi} (z,\mu)}{\partial z} + (1-\mu^{2})^{\frac{1}{2}} {\bf{A}} \ \vec{\psi}(z,\mu)  \\
= \frac{2c}{\pi} (1-\mu^{2})^\frac{1}{2} \ {\bf{B}} \ \int_{-1}^{1} (1-\mu'^{2})^\frac{1}{2} \ \vec{\psi}(z,\mu') \,d\mu' ,
\end{multline}
where $\vec{\psi} = [\psi_1, \psi_2, \dots, \psi_N]^\mathrm{t}.$

The boundary conditions are given as column vectors, 
\begin{equation}
\vec{\psi}(0,\mu) = {\bf{F(\mu)}} \quad \vec{\psi}(Z,-\mu) = {\bf{G(\mu)}},
\end{equation}  
where the components $\psi_{i}$ are expressed as
\begin{equation}
{F}_{i}(\mu) = \frac{1}{2\pi A} \int_{R}\int_{0}^{2\pi} \beta_{i} f \, d\phi\, dx \,dy , \ \mu > 0, 
\end{equation}
\begin{equation}
{G}_{i}(\mu) = \frac{1}{2\pi A} \int_{R}\int_{0}^{2\pi} \beta_{i} g \, d\phi \,dx\, dy , \ \mu < 0 .
\end{equation}
${\bf{A}}$ and ${\bf{B}}$ are $ N \times N$ matrices consisting of elements $a_{ij},b_{ij}$, where
\begin{equation}
a_{ij} = \frac{1}{2 \pi A} \left[ \ \int_{\partial R} \int_{\omega \cdot n>0 } \vec{\omega} \cdot \vec{n} \  \beta_{i}  \alpha_{j}  \,d \phi\, ds  - \int_{R} \int_{0}^{2 \pi} (\vec{\omega} \cdot \nabla \beta_{i}) \ \alpha_{j} \,d \phi \,ds \ \right],
\end{equation}

\begin{equation}
b_{ij} = \frac{1}{4 \pi A} \int_{\partial R} \left[\left( \int_{\omega \cdot n < 0} \vec{\omega} \cdot \vec{n} \right) \beta_{i} \,d \phi \ \times \left( \int_{\vec{\omega} \cdot \vec{n} > 0} \vec{\omega} \cdot \vec{n} \ \alpha_{j} d\phi\right)\right]\, ds.
\end{equation}

\subsection{Basis Functions}

The basis functions for the $N=3$ model are derived independently by Larsen et al. (1986) and Garcia (2000). In each case, a Galerkin scheme  was used to select the corresponding weight functions, i.e. $\beta_{i}(x,y,\phi) = \alpha_{i}(x,y,\phi)$ for $i=1,2,3$. The first and second basis functions are expressed as \cite{Larsen:1983uh}
\begin{equation}
\alpha_{1}(x,y,\phi) = 1,
\label{b1}
\end{equation}
\begin{equation}
\alpha_{2}(x,y,\phi) = u[D(x,y,\vec{\omega}) - v],
\label{b2}
\end{equation}	
where $u$ and $v$ are constants needed to satisfy the orthonormal condition in Eq.~\eqref{eq:normal}. The first two basis functions are linear combinations of 1 and $D(x,y,\omega)$, where $D$ defines the distance from a point $(x,y,z)$ in the duct's interior to it's inner wall $\partial R$ along the direction $-\omega$. The constants $u$ and $v$ are given as  	
\begin{equation}
u = \left( \frac{1}{2\pi}  \int_{R} \int_{0}^{2 \pi} [D(x,y,\vec{\omega}) - v]^{2} \, d\phi \,d x \,d y \right)^\frac{-1}{2},
\label{u} 
\end{equation}
\begin{equation}
v =  \frac{1}{2\pi}  \int_{R} \int_{0}^{2 \pi} D(x,y,\vec{\omega})  \, d\phi \,d x \,d y.   
\label{v}
\end{equation}	
The third basis function is expressed as a linear combination of 1, $D(x,y,\vec{\omega})$, and $D^{2}(x,y,\vec{\omega})$ \cite{Garcia:2000bw}:
\begin{equation}
\alpha_{3}(x,y,\phi)=r[D(x,y,\vec{\omega} - v)][D(x,y,\omega) - v - q] - \frac{r}{u^{2}},
\label{b3}
\end{equation}
where the constant $q$ is defined as 
\begin{equation}
q = \frac{u^{2}}{2\pi} \int_{R} \int_{0}^{2\pi} [D(x,y,\vec{\omega}) - v]^{3} \,d\phi \,d x \,d y,
\label{q}
\end{equation} 
and $r$ is defined as 
\begin{equation}
r =  \left[ \frac{1}{2 \pi} \int_{R} \int_{0}^{2 \pi} [D(x,y,\vec{\omega}) - v]^{4} \, d\phi\,d x \,d y  - \frac{(q^{2} + \frac{1}{u^{2}})}{u^{2}} \right]^{\frac{1}{2}}.
\label{r} 
\end{equation}

As previously noted, the basis and weights functions are not directly expressed in the multi-basis function form of the transport equation. Rather, these functions compose the elements of the matrices $\bf{A}$ and $\bf{B}$, which are required to apply the model. Having explicit statements of the first thee basis functions, as well as their dependencies (constants: $u,v,q,r$), exact expressions for the matrix elements can be found by direct substitution into Eqs.~(\ref{b1}), (\ref{b2}), and (\ref{b3}). The matrix elements for the $N=3$ model in a circular duct are given in \cite{Garcia:2000bw}.

\section{Wall Migration Model: Methods and Parameters} 
In this section we will develop a nonlocal transport equation that accounts for particle migration into duct walls, composed of a specified material, using the third-order approximate one-dimensional model. Later, this nonlocal equation is specifically applied to ducts with iron, concrete, and graphite walls, subject to a thermal neutron source.   

In order to account for neutral particles that migrate a fixed distance in the walls and undergo diffuse emission, Prinja \cite{Prinja:1996kl} introduced a nonlocal kernel density $K(z' \rightarrow z)$ that gives the probability density of a particle striking the wall at $z'$ re-entering the duct at point $z$. He this kernel applied for the $N=1$ case for a semi-infinite duct. The kernel density satisfies the condition 
\begin{equation}
\int_{-\infty}^{\infty} K(z' \rightarrow z) \,dz= 1,
\end{equation}   
where this ensures that if the ducts is infinite in length the particle is must be re-emitted somewhere. For the form of this kernel, Prinja \cite{Prinja:1996kl} proposed an exponential function expressed as
\begin{equation}
K(z' \rightarrow z) = \frac{\lambda}{2} \exp(-\lambda |z-z'|),
\end{equation}
where $\lambda$ is a free parameter where $\lambda^{-1}$ is the average net-distance traveled between entering the wall and re-emerging in the duct. Additionally, 
\begin{equation}
\lim_{\lambda\to\infty} K(z' \rightarrow z) = \delta(z-z'), 
\end{equation}
ensures recovery of the local model. A full account of wall migration requires integrating the kernel density over the entire length of the duct, where
\begin{equation}
\int_{0}^{Z} K(|z-z'|) dz' = \int_{0}^{Z} \frac{\lambda}{2} \exp(-\lambda|z-z'|),  
\end{equation}  
is introduced into the scattering term of the transport equation. Accounting for wall migration via the kernel density, the approximate one-dimensional duct model transport equation is expressed as
\begin{multline}
\mu \frac{\partial \vec{\psi} (z,\mu)}{\partial z} + (1-\mu^{2})^{\frac{1}{2}} {\bf{A}} \ \vec{\psi}(z,\mu) = \frac{2c}{\pi} (1-\mu^{2})^\frac{1}{2} \\ \times {\bf{B}} \ \int_{0}^{Z} \partial z'\int_{-1}^{1} (1-\mu'^{2})^\frac{1}{2} \ K(z' \rightarrow z) \ \vec{\psi}(z,\mu') \,d\mu',  
\label{1dmc}
\end{multline}
where $Z$ is the length of the duct. 

\subsection{Estimating the parameters}
From Eq.~(\ref{1dmc}) it is clear that in addition to the basis and weight functions, the parameters $\lambda$ and $c$ must be known in order to apply the model accounting for wall migration. 
As noted, the parameter $c$ defines the probability that a particle will re-enter the duct. 

These parameters, $\lambda$ and $c$ need to be estimated and will be a function of the material, the thickness of the duct walls, and the particle type. Here, we will focus on thermal neutrons as the particles of interest, though different energies or particles could be treated using our prescription. To estimate $c$ and $\lambda$  we simulate a point source  of thermal neutrons with a cosine distribution on the surface of a disk 20 cm thick with a radius of 100 cm using MCNP \cite{MCNP}.  The value of $c$ is computed as the ratio of the current entering the disk to the current exiting the disk. This calculation is similar to that used in \cite{Garcia:2003fl}. 
The value of $c$ for thermal neutrons is given in Table \ref{tab:c} for each of three materials: natural iron, ordinary concrete, and graphite. The material compositions are defined according to PNNL Compendium of Material Composition for Radiation Transport Modeling \cite{Williams:2006}. 
\begin{table}[H]
	\centering
	\caption{ Values for wall reflection probability computed using MCNP for $10^6$ histories.}\label{tab:c}
	\begin{tabular}{cc}
		\hline
		{Material} & {$c$}  \\\hline
		Iron &  0.54   \\  
		Concrete &  0.70 \\  
		Graphite &  0.85 \\ 
		\hline		
	\end{tabular}
\end{table}      

Using MCNP's PTRAC function, which follows the life of individual particles including terminal events, the mean radial distance that particles travel in the disk before emerging on the side they entered can be calculated. This average distance is then interpreted as $\lambda^{-1}$ 

\vspace{0.7cm}

\begin{figure}[H]
	\centerline{\includegraphics[width=\textwidth]{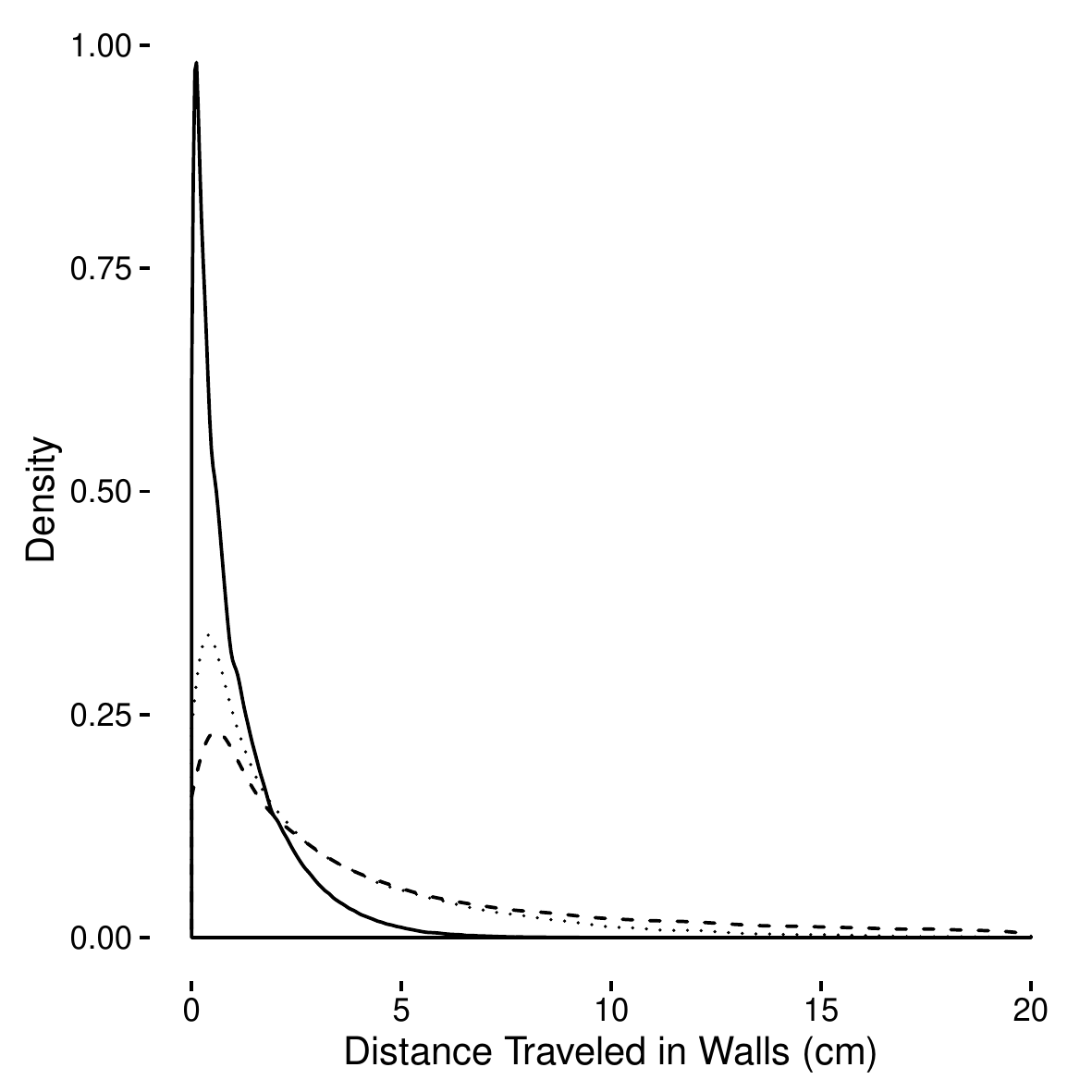}}
	\caption{Density plot of the distances traveled by particles in the disk for different materials. Iron (solid) has the narrowest distribution, with a mean of approximately 1 cm, followed by concrete (dotted) with a mean of approximately 2.8 cm, and graphite (dash) with a mean of approximately 5.9 cm. }\label{fig:exp}
\end{figure}

\begin{table}[H]
	\centering
	\caption{Mean distance $\lambda^{-1}$ (cm) traveled in walls computed using MCNP for $10^6$ histories, and material density $\rho$ (g/$cm^{3}$).}
	\label{table:d}
	\begin{tabular}{ccc}
		\hline
		{Materials} &  ${\lambda^{-1}}$ & {$\rho$}  \\\hline
		Iron  & 1.07 &  7.87 \\ 
		Concrete & 2.80 &  2.30 \\ 
		Graphite & 5.91  &  1.70 \\
		\hline
	\end{tabular}
\end{table}     
Figure \ref{fig:exp} shows the distribution of radial distances traveled in the walls by particles, for all three materials.  The shape of the distributions figure indicates that the exponential model for wall migration is a reasonable approximation.
Given the wall reflection constant and the kernel density, as well as the matrix elements of $\bf{A}$ and $\bf{B}$, the $N=3$ nonlocal transport equation can be fully solved using a discrete ordinates method.

\section{Circular Duct Results}
We solve Eq.~\eqref{1dmc} using the discrete ordinates method with the diamond difference discretization in space and {160 mesh points} and {640 angles}.  The spatial integration in the scattering kernel is approximated using a Gauss-Kronrod quadrature rule.

%

As a test-case, the $N=3$ nonlocal, approximate one-dimensional model is applied to a duct of circular cross section with varying radius $\rho$. The walls of the duct are 20 cm thick (corresponding to the thickness of the disk used to calculate $c$ and $\lambda^{-1}$), and the duct is 100 cm long. A second test-case is considered using the same duct geometry with a fixed 10 cm radius, 20 cm thick walls, and variying length. The quantities of interest in this case are reflection probability and transmission probability, corresponding to natural iron, ordinary concrete, and graphite ducts, subject to a thermal neutron source. The reflection ($R$) and transmission ($T$) probabilities are calculated as
\begin{equation}
R = 2 \int_{0}^{1} = \mu \Psi_{1}(0,-\mu) d \mu,
\end{equation}
\begin{equation}
T = 2 \int_{0}^{1} = \mu \Psi_{1}(Z,\mu) d \mu,
\end{equation}
for values of $c$ and $\lambda$ corresponding to each of the wall materials.  


The results for Tables \ref{table:r3}-\ref{table:t14} correspond to a Gauss-Legendre quadrature of 640 angles and weights, coupled with a spatial discretization mesh consisting of 160 intervals. Reflection and transmission probabilities are calculated using Eq.~\eqref{1dmc} and compared to MCNP simulations for $10^{6}$ histories. In each table the nonlocal one-dimensional transport model is abbreviated as ``1DWM", short for 1-D wall migration. The percent deviation ($ \%$ Dev) is calculated as 
\vspace{0.5cm}
\begin{equation}
\% \ \mbox{Deviation} = \frac{P_{1DWM} - P_{MCNP}}{P_{MCNP}} \times 100,
\end{equation}
where $P$ is the reflection or transmission probability.

\vspace{0.7cm}

\begin{table}[H] 
	\centering
	\caption{ Reflection probability for iron ducts of 100 cm length, 20 cm wall thickness, and varying radius.} 
		\label{table:r3}
	\begin{tabular}{cccc}
		\hline 
		{Radius}& {1DWM} & {MCNP}& {\% Dev.} \\ \hline 
		8&	0.1557&	0.1466&	6.23\\ 
		10&	0.1597&	0.1529&	4.48\\  
		15&	0.1653&	0.1621&	2.03\\ 
		20&	0.1679&	0.1665&	0.84\\ 
		30&	0.1691&	0.1693&	-0.10\\ 
		50&	0.1630&	0.1638&	-0.49\\ 
		\hline
	\end{tabular}
\end{table}

\begin{table}[H]
	\centering
	\caption{ Transmission probability for iron ducts of 100 cm length, 20 cm wall thickness, and varying radius.}
	\label{table:t4}
	\begin{tabular}{cccc}
		\hline
		{Radius}& {1DWM} & {MCNP}& {\% Dev.}\\
		\hline
		8&  0.007950&	0.008743&	-9.98\\ 
		10& 0.01340&	0.014736&	-9.95\\ 
		20&	0.03381&	0.03716&	-9.91\\ 
		20&	0.06262&	0.06778&	-8.23\\ 
		30&	0.1322&	0.1390&	-5.12\\ 
		50&	0.2691&	0.2751&	-2.26\\ \hline
		
	\end{tabular}
\end{table}

Tables \ref{table:r3}-\ref{table:t4} show the results for thermal neutrons transported in iron ducts, and indicate that the 1DWM produces relatively low errors for both the reflection and transmission probability values as functions of radius, giving mean percent deviations of 2.17\% and -7.58\%. In general the error decreases as the radius increases, but the model consistently overestimates the reflection probability and underestimates the transmission probability. These trends hold for all three materials, however, the 1DWM produces notably higher errors for graphite and concrete. Tables \ref{table:r7}-\ref{table:t12} show that the 1DWM produces mean percent deviations for the reflection and transmission probability as a function of radius of 13.57\% and -17.43\% for concrete and 15.56\% and -14.85\% for graphite.

\begin{table}[H] 
	\centering
	\caption{Reflection probability for concrete ducts of 100 cm length, 20 cm wall thickness, and varying radius.} 
	\label{table:r7}
	\begin{tabular}{cccc}
		\hline 
		{Radius}& {1DWM} & {MCNP}& {\% Dev.}\\
		\hline
		8&	0.2017&	0.1664&	21.22\\ 	
		10&	0.2139&	0.1812&	17.92\\ 	
		15&	0.2326&	0.2050&	13.41\\ 	
		20&	0.2419&	0.2174&	11.27\\ 	
		30&	0.2473&	0.2263&	9.28 \\ 	
		50&	0.2362&	0.2180&	8.31 \\ \hline 	
		
	\end{tabular}
\end{table}

\begin{table}[H] 
	\centering
	\caption{Transmission probability for concrete ducts of 100 cm length, 20 cm wall thickness, and varying radius.} 
	\label{table:t8}
	\begin{tabular}{cccc}
		\hline
		{Radius}& {1DWM} & {MCNP}& {\% Dev.}\\
		\hline
		8&  0.009983&	0.01314&	-24.02\\ 	
		10& 0.01799&	0.02355&	-23.60\\ 	
		15& 0.04778&	0.0609&	-21.48\\ 	
		20& 0.08730&	0.1059&	-17.52\\ 	
		30& 0.1734&	0.1972&	-12.01\\ 	
		50&	0.3228&	0.3453&	-6.51\\ \hline	
		
	\end{tabular}	
\end{table}

\begin{table}[H] 
	\centering
	\caption{Reflection probability for graphite ducts of 100 cm length, 20 cm wall thickness, and varying radius.}
	\label{table:r11}
	\begin{tabular}{cccc}
		\hline 
		{Radius}& {1WMD} & {MCNP}& {\% Dev.}\\
		\hline
	8	&0.2216&	0.1794&	23.55 \\
	10	&0.2479&	0.2031&	22.05\\
	15	&0.2879&	0.2452&	17.40\\
	20	&0.3083&	0.2705&	13.96\\
	30	&0.3203&	0.2910&	10.06\\
	50	&0.3023&	0.2842&	6.36	\\ \hline
	\end{tabular} 
\end{table}	
\begin{table}[H] \small
	\centering
	\caption{ Transmission probability for graphite ducts of 100 cm length, 20 cm wall thickness, and varying radius.}
	\label{table:t12}
	\begin{tabular}{cccc}
		\hline 
		{Radius}& {1DWM} & {MCNP}& {\% Dev.}\\
		\hline	
	8	&0.01931&	0.01414&	-36.60 \\
	10	&0.03315&	0.02621&	-26.49\\
	15	&0.07872&	0.06915&	-13.85\\
	20	&0.1307&	0.1210&	-8.01\\
	30	&0.2289&	0.2213&	-3.44\\
	50	&0.3801&	0.3761&	-1.06\\
	 \hline
		
	\end{tabular} 
\end{table}

The 1DWM results can be generally compared to the results produced by Garcia's local albedo model \cite{Garcia:2003fl}, where the same duct configurations and radial values are applied, but the compostion of concrete differs and Iron-56 is used instead of natural iron (graphite is not included). For thermal neutrons in Iron-56 and concrete ducts, Garcia \cite{Garcia:2003fl} reports mean percent deviations of 11.60\% and 56.69\% for the reflection probability, and 5.16\% and 6.21\% for the transmission probability. The 1DWM performs significantly better for the reflection probabilities, overcoming the local albedo model's edge effects (i.e in the abscence of migration, particles interacting at the edge of the duct will be reflected out without interacting a second time). For transmission probability, however, the local albedo model performs better due to the multigroup approach. As the 1DWM does not account for energy dependence, information is lost as neutrons transport across the wall materials ; this effect is more poignantly seen for concrete and graphite which have significantly smaller absorption cross-sections and larger scattering cross-sections than natural iron.

\begin{table}[H] 
	\centering
	\caption{ Reflection probability for iron ducts with a 10 cm radius, 20 cm wall thickness, and varying  length.} 
	\label{table:r5}
	\begin{tabular}{cccc}
		\hline 
		{Length}& {1DWM} & {MCNP}& {\% Dev.} \\ \hline 
		10&0.1166	&0.1081	&7.91	\\ 
		30&	0.1562& 0.1486 	&5.06	\\  
		50&0.1592	&0.1523	& 4.53	\\ 
		70&0.1596	&0.1528	& 4.46	\\ 
		150&0.1601 &0.1529 & 4.71	\\ 
		200&0.1606 &0.1529 & 5.04	\\ 
		\hline
	\end{tabular}
\end{table}

\begin{table}[H]
	\centering
	\caption{ Transmission probability for iron ducts with a 10 cm radius, 20 cm wall thickness, and varying length.}
	\label{table:t6}
	\begin{tabular}{cccc}
		\hline
		{Length}& {1DWM} & {MCNP}& {\% Dev.} \\ \hline 
		10&	0.4846& 0.4818 & -0.59	\\ 
		30& 0.1541	& 0.1594  & -3.32	\\  
		50& 0.06217	& 0.06695	& -7.15	\\ 
		70& 0.03018	& 0.03317	& -9.03\\ 
		150& 0.005278 & 0.005749  & -8.20 	\\ 
		200& 0.002813 & 0.002988 & -5.86  \\  \hline
		
	\end{tabular}	
\end{table}

\begin{table}[H] 
	\centering
	\caption{ Reflection probability for concrete ducts with a 10 cm radius, 20 cm wall thickness, and varying length.} 
	\label{table:r9}
	\begin{tabular}{cccc}
		\hline 
		{Length}& {1DWM} & {MCNP}& {\% Dev.} \\ \hline 
		10& 0.1256	&0.1015	& 23.74	\\ 
		30&	 0.2028 & 0.1694 	& 19.72	\\  
		50&	0.2123 & 0.1790	& 18.60	\\ 
		70&	0.2139 & 0.1808	& 18.30	\\ 
		150& 0.2135 & 0.1814 & 17.68 	\\ 
		200& 0.2147 & 0.1814  & 18.38\\ 
		\hline
	\end{tabular}
\end{table}
\begin{table}[H]
	\centering
	\caption{ Transmission probability for concrete ducts with a 10 cm radius, 20 cm wall thickness, and varying length.}
	\label{table:t10}
	\begin{tabular}{cccc}	
		\hline
		{Length}& {1DWM} & {MCNP}& {\% Dev.} \\ \hline 
		10& 0.4975	& 0.4994  & -0.38 	\\ 
		30& 0.1923	& 0.2074  & -7.28	\\  
		50& 0.08599  & 0.1007 	& -14.56	\\ 
		70& 0.04277	& 0.05336	& -19.84	\\ 
		150& 0.006384 & 0.008264  & -22.76	\\ 
		200& 0.003169 & 0.003894  & -18.61  \\  \hline
	\end{tabular}
\end{table}

\begin{table}[H] 
	\centering
	\caption{ Reflection probability for graphite ducts with a 10 cm radius, 20 cm wall thickness, and varying length.} 
	\label{table:r13}
	\begin{tabular}{cccc}
		\hline 
		{Length}& {1DWM} & {MCNP}& {\% Dev.} \\ \hline 
	10	&0.1033&	0.09515& 8.59 \\
	30	&0.2116&	0.1796&	17.81\\
	50	&0.2389&	0.1976&	20.85\\
	70	&0.2465&	0.2019&	22.11\\
	150	&0.2502&	0.2033&	23.07\\
	200	&0.2503&	0.2034&	23.06\\ 	
		\hline
	\end{tabular}
\end{table}
\begin{table}[H]
	\centering
	\caption{ Transmission probability for graphite ducts with a 10 cm radius, 20 cm wall thickness, and varying length.}
	\label{table:t14}
	\begin{tabular}{cccc}	
		\hline
		{Length}& {1DWM} & {MCNP}& {\% Dev.} \\ \hline 
	10	&0.4797&	0.4765&	-0.68 \\
	30	&0.2230&	0.2034&	-9.67 \\
	50	&0.1235&	0.1053&	-17.21\\
	70	&0.07152&	0.05803&	-23.26\\
	150	&0.01125&	0.008734&	-28.79\\
	200	&0.004754&	0.003941&	-20.62
		 \\  \hline
	\end{tabular}
\end{table}

Tables \ref{table:r5} - \ref{table:t14} show that the 1DWM performs similarly across wall materials for length dependence, as it does for radial dependence. The mean percent deviations for the reflection and transmission probability are 5.33\% and -5.69\% for natural iron, 19.40\% and -13.90 for concrete, and 19.78\% and -16.705\% for graphite. For each of the materials both the 1DWM and the MCNP results indicate that as the duct length increases--- at approximately 100-150 cm--- both the reflection and transmission probability have plateaued; as such, we see slight decreases in error at a length of 200 cm.  

A notable trend in the length-dependent results is that the 1DWM performs very well for 10 cm ducts, with the exception of the reflection probability for concrete. In very short ducts, assuming monoenergetic particles has a less significant impact, particularly near the duct entrance where the particles are largely of uniform energy--- which also explains why the 1DWM provides more accurate values for radially dependent reflection probabilities than does the local albedo model. Concrete is generally difficult to model using a monoenergetic assumption---even for short ducts--- given that it contains very light elements, resulting in a higher average energy loss per collision.

\begin{figure}
	\centerline{\includegraphics[width=1.1\textwidth]{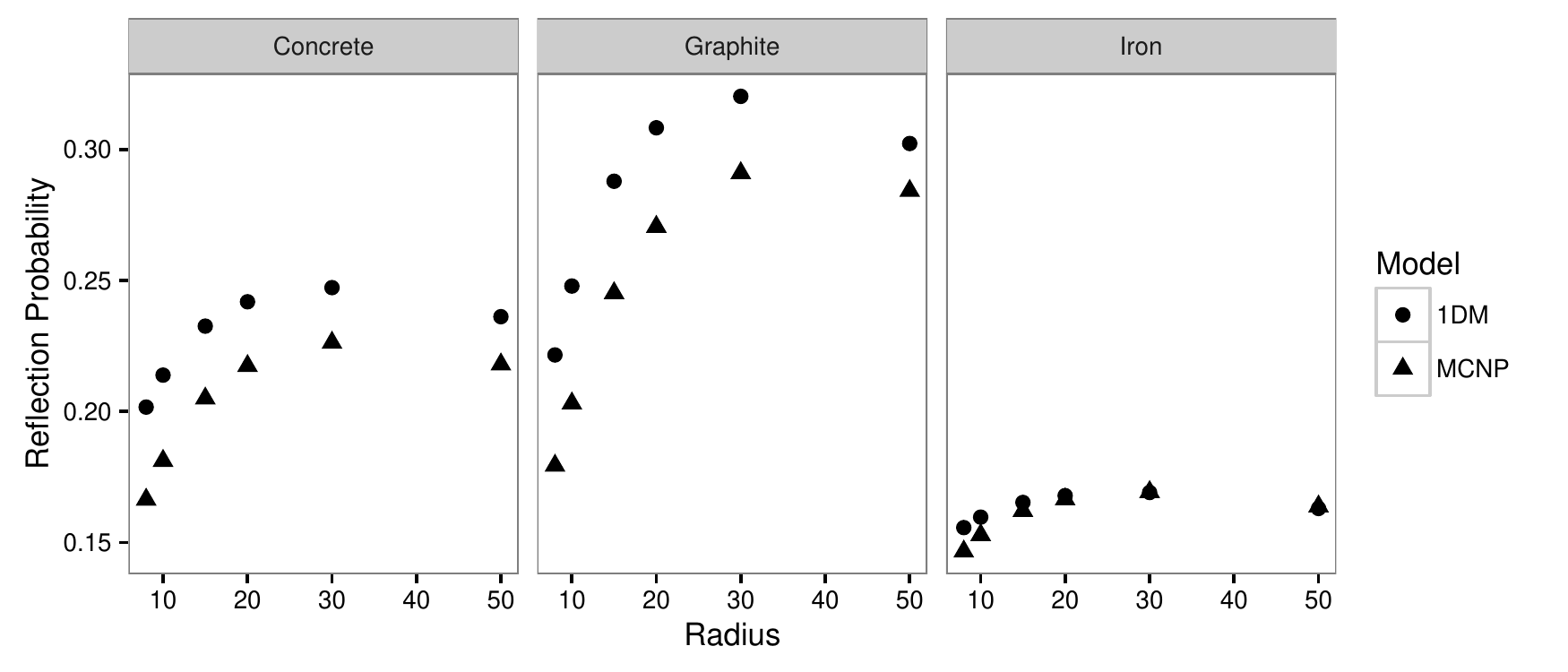}}
	\label{fig:exp1}
	\caption{Reflection probability values, with respect to radius, calculated by the 1DWM versus MCNP.}
\end{figure} 
\begin{figure}
	\centerline{\includegraphics[width=1.1\textwidth]{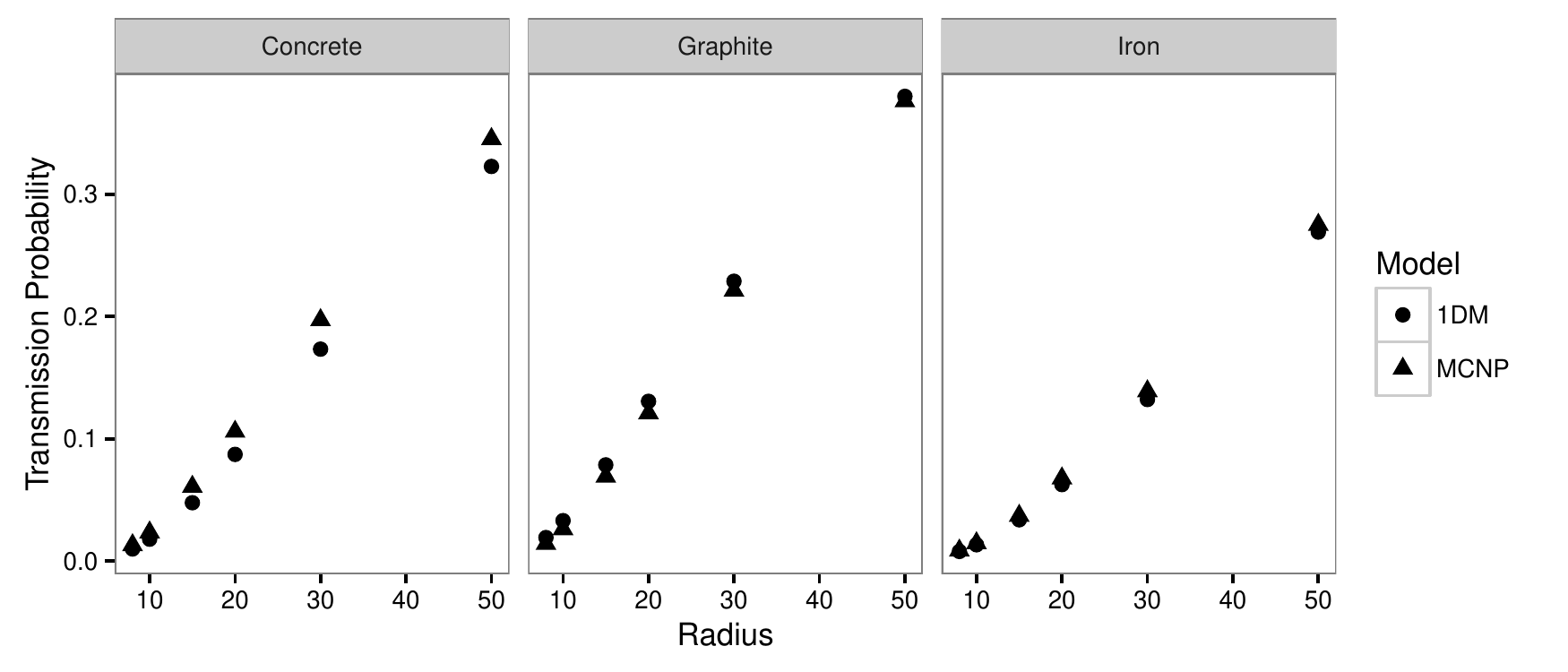}}
	\label{fig:exp2}
	\caption{Transmission probability values, with respect to radius, calculated by the 1DWM versus MCNP.}
\end{figure}

\begin{figure}
	\centerline{\includegraphics[width=1.1\textwidth]{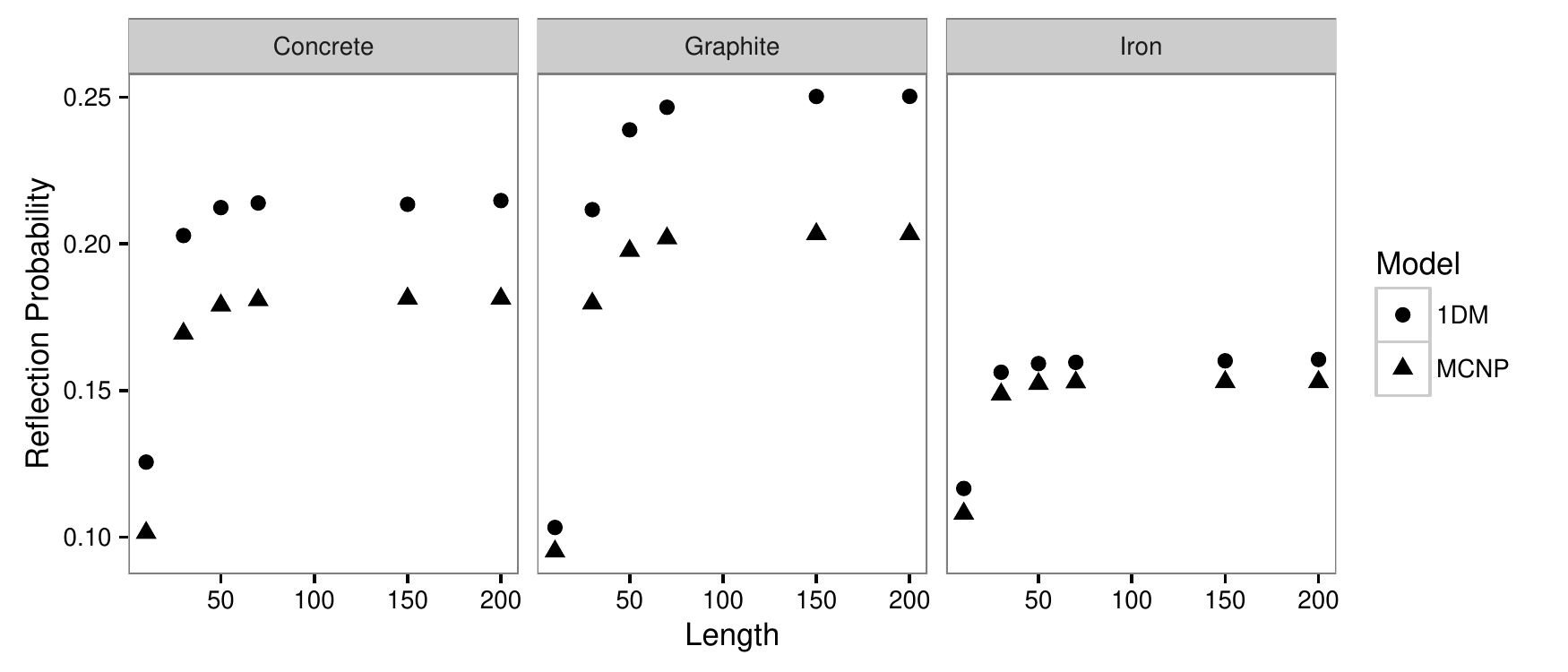}}
	\label{fig:exp3}
	\caption{Reflection probability values, with respect to length, calculated by the 1DWM versus MCNP.}
\end{figure} 
\begin{figure}
	\centerline{\includegraphics[width=1.1\textwidth]{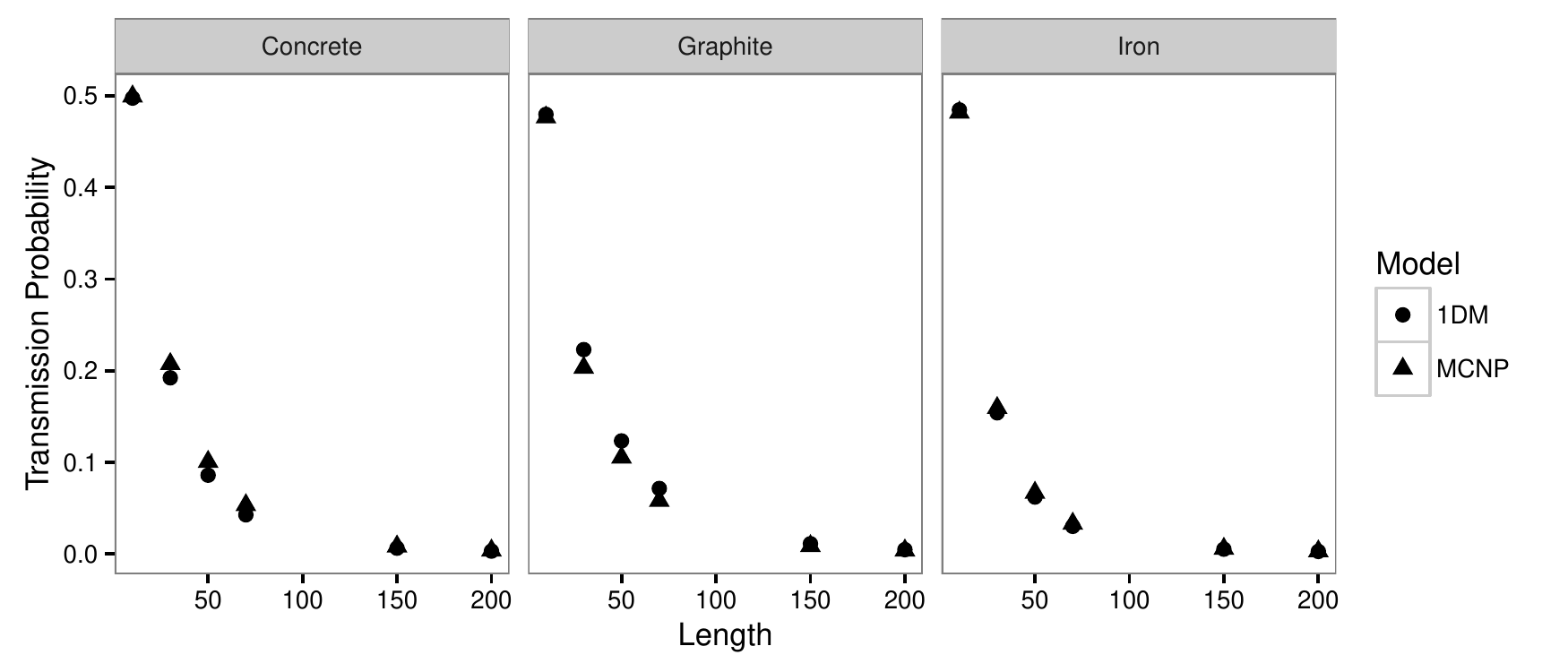}}
	\label{fig:exp4}
	\caption{Transmission probability values, with respect to length, calculated by the 1DWM versus MCNP.}
\end{figure}
\vspace{15cm}

\section{Concluding Remarks}

The problem of  neutral particle transport in a duct is treated using an approximate one-dimensional model for the third basis function, accounting for wall migration via a scattering kernel density. Comparisons to MCNP results demonstrate that the $N=3$ nonlocal approximate one-dimensional model performs well for iron ducts, while producing higher levels of error for concrete and graphite ducts. The higher error in concrete and graphite ducts, may be attributed to a greater energy dependence, due to their relatively large scattering cross-section and small absorption cross-section. 

Garcia \cite{Garcia:2003fl} addresses the problem of energy-dependence by introducing a multi-group albedo approximation, where a value for wall reflection probability ($c$) is calculated for a number of subgroups in both the thermal and fast-range. A natural progression of this work would be the introduction of a scattering kernel density, accounting for the distance traveled by particles in the walls, corresponding to specified energy groups.   

\bibliographystyle{unsrt}
\bibliography{references}

\end{document}